\documentclass[structabstract]{aa}

\usepackage{graphicx}
\usepackage{rotating}
\usepackage{txfonts}

\def\dsct{$\delta$~Scuti }
\def\gdor{$\gamma$~Doradus }
\def\Msun{$M_{\odot}$}
\def\Lsun{$L_{\odot}$}
\def\Rsun{$R_{\odot}$}

\def\Teff{\ensuremath{T_{\mathrm{eff}}}}
\def\cd{d$^{\rm -1}$}
\def\logg{\ensuremath{\log g}}
\def\vmic{$\upsilon_{\mathrm{mic}}$}
\def\vmac{$\upsilon_{\mathrm{mac}}$}
\def\vsini{\ensuremath{{\upsilon}\sin i}}
\def\kms{$\mathrm{km\,s}^{-1}$}

\def\vrad{${\upsilon}_{\mathrm{r}}$}
\def\llm{{\sc LLmodels}}

\def\vald{{\sc VALD}}
\def\synth{{\sc SYNTH3}}
\def\logl{\ensuremath{\log L/L_{\odot}}}

\begin{document}

\title{\gdor pulsation in two pre-main sequence stars discovered by CoRoT \thanks{The CoRoT  space mission was developed and is operated by the French space agency CNES, with participation of ESA's RSSD and Science Programmes, Austria, Belgium, Brazil, Germany, and Spain.}}

\author{K. Zwintz\inst{1,2}\thanks{Pegasus Marie Curie post-doctoral fellow of the Research Foundation - Flanders} \and
L. Fossati\inst{3} \and
T. Ryabchikova\inst{4} \and
A. Kaiser\inst{1} \and
M. Gruberbauer\inst{5} \and
T. G. Barnes\inst{6}  \and
A. Baglin\inst{7}  \and
S. Chaintreuil\inst{7}
}

\offprints{K. Zwintz, \\ \email{konstanze.zwintz@univie.ac.at}}

\institute{
   University of Vienna, Institute of Astronomy, T\"urkenschanzstrasse 17, A-1180 Vienna, Austria \\
    \email konstanze.zwintz@univie.ac.at  \and
    Instituut voor Sterrenkunde, K. U. Leuven, Celestijnenlaan 200D, B-3001 Leuven, Belgium \and
   Argelander-Institut f\"ur Astronomie der Universit\"at Bonn, Auf dem H\"ugel 71, 53121 Bonn, Germany \and
   Institute of Astronomy, Russian Academy of Sciences, Pyatnitskaya 48, 119017 Moscow, Russia \and
   Department of Astronomy and Physics, St. MaryÕs University, Halifax, NS B3H 3C3, Canada \and
      The University of Texas at Austin, McDonald Observatory, 82 Mt. Locke Rd., McDonald Observatory, Texas 79734, USA  \and
       LESIA, Observatoire de Paris-Meudon, 5 place Jules Janssen, 92195 Meudon, France
    }

\date{Received / Accepted }

\abstract
{Pulsations in pre-main sequence stars have been discovered several times within the last years. But nearly all of these pulsators are of $\delta$~Scuti-type. $\gamma$~Doradus-type pulsation in young stars has been predicted by theory, but lack observational evidence. }
{We present the investigation of variability caused by rotation and ($\gamma$~Doradus-type) pulsation in two pre-main sequence members of the young open cluster NGC\,2264 using high-precision time series photometry from the CoRoT satellite and dedicated high-resolution spectroscopy.}
{The variability found using the CoRoT data was combined with the fundamental parameters and chemical abundances derived from high-resolution spectroscopy, obtained at the Mc Donald Observatory, to discuss the presence of pulsation and rotation in the two NGC\,2264 cluster members.
Time series photometry of NGC\,2264\,VAS\,20 and NGC\,2264\,VAS\,87 was obtained by the CoRoT satellite during the dedicated short run SRa01 in March 2008. NGC\,2264\,VAS\,87 was re-observed by CoRoT during the short run SRa05 in December 2011 and January 2012.  Frequency analysis was conducted using Period04 and SigSpec. The spectral analysis was performed using equivalent widths and spectral synthesis. }
{The frequency analysis yielded ten and fourteen intrinsic frequencies for NGC\,2264\,VAS\,20 and NGC\,2264\,VAS\,87, respectively, in the range from 0 to 1.5\cd which are attributed to be caused by a combination of rotation and pulsation. The effective temperatures were derived to be 6380$\pm$150\,K for NGC\,2264\,VAS\,20 and 6220$\pm$150\,K for NGC\,2264\,VAS\,87. Membership of the two stars to the cluster is confirmed independently using X-ray fluxes, radial velocity  measurements and proper motions available in the literature. The derived Lithium abundances of $\log\,n$(Li) = 3.34 and 3.54 for NGC\,2264\,VAS\,20 and NGC\,2264\,VAS\,87, respectively, are in agreement with the Lithium abundance for other stars in NGC\,2264 of similar \Teff\ reported in the literature. }
{We conclude that the two objects are members of NGC 2264 and therefore are in their pre-main sequence evolutionary stage.  Their variability is attributed to be caused by rotation and $g$-mode pulsation rather than rotation only. Assuming that part of their variability is caused by pulsation, these two stars might be the first pre-main sequence $\gamma$ Doradus candidates. }

\keywords{stars: pre-main sequence - stars: oscillations - stars: individual: NGC 2264 VAS 20, NGC 2264 VAS 87 - techniques: photometric - techniques: spectroscopic}

\titlerunning{PMS \gdor\ candidates in NGC\,2264}
\authorrunning{K. Zwintz et al.}
\maketitle

\section{Introduction}

Pre- and post-main sequence evolutionary tracks intersect close to the zero-age main sequence (ZAMS) and cross the instability regions of \dsct\ and \gdor\ stars. Stars of both evolutionary stages are therefore expected to show both types of pulsations. \dsct\ and \gdor\ pulsators are well established among the hydrogen core burning (post-) main sequence stars (e.g., Rodriguez et al. \cite{rod00}, Handler \cite{han99}).
Pulsations in pre-main sequence (PMS) stars, which gain their energy mainly from gravitational contraction, have been reported several times within the last years (e.g., Ripepi et al. \cite{rip06}, Zwintz et al. \cite{zwi09}), but the majority of these pulsators are of \dsct-type. 
PMS \dsct\ stars pulsate in $p$-modes and have spectral types from A to early F; their pulsation periods range from $\sim$ 20 minutes to 6.5 hours.
However, recently a PMS \dsct\--\gdor\ hybrid star, CoID0102699796, was found using data obtained with the CoRoT satellite (Ripepi et al. \cite{rip11}).

The classical (post-) main sequence \gdor\ stars have spectral types from late A to F and are known to pulsate in high order $g$-modes with periods ranging from approximately 0.3 to a few days (e.g., Kaye et al. \cite{kay99}). The modulation of the radiative flux at the base of the convective envelope has been suggested to be the excitation mechanism for \gdor\ stars (e.g. Guzik et al, \cite{guz00}, Dupret et al. \cite{dup05}). 
Similar properties are expected for \gdor\ stars in their PMS evolutionary phase. Bouabid et al. (\cite{bou11}) determined the theoretical seismic properties of PMS \gdor\ pulsators, but no confirmed members of this group have been reported yet.

The search for \gdor\ pulsation in PMS stars is complicated by the likely presence of rotational modulation due to spots in such young objects. Candidates for PMS \gdor\ pulsation are located in the same region of the Hertzsprung Russell (HR) diagram as the hotter members  of the group of T Tauri stars. The light curves of T Tauri stars are well-known to show regular and irregular variations caused by spots on the surface and the interaction of the stars with their circumstellar environment (e.g., Alencar et al. \cite{ale10}).
\gdor\ type pulsation occurs on the same time scales as variability caused by spots and surface inhomogeneities, i.e., with periods on the order of up to few days. It is therefore hard to unambiguously distinguish between the influence of rotation and pulsation in such stars (see, e.g., Uytterhoeven et al. \cite{uyt10}).

As \gdor\ stars have periods on the order of up to a few days, they are not easy to observe from ground. Only with dedicated multi-site campaigns it is possible to resolve these frequencies, both spectroscopically and photometrically. The high-precision photometry obtained with the CoRoT satellite allowed for the first time to study these stars in more detail. 

In this study we present the analysis of CoRoT light curves and high-resolution spectroscopy for two PMS members of the young open cluster NGC 2264.

\section{NGC\,2264: VAS\,20 and VAS\,87}

The young open cluster NGC\,2264 ($\alpha_{2000}$ = $6^h$ $41^m$, $\delta_{2000}$ = $+9^{\circ}$ $53'$) has been studied frequently in all wavelength regimes in the past. It is located in the Monoceros OB1 association about 30\,pc above the galactic plane and has a diameter of $\sim$39 arcminutes. Sung et al. (\cite{sun97}) report a cluster distance of 759 $\pm$ 83pc which corresponds to a distance modulus of 9.40 $\pm$ 0.25 mag. 

The age of NGC\,2264 can only be determined with a relatively large error as its main sequence consists only of massive O and B stars and stars of later spectral types are still in their pre-main sequence phase. Additionally, not all cluster members might have formed at the same time and star formation is reported to still be going on in the cluster (e.g., Flaccomio et al. \cite{fla99}). Therefore different values for NGC\,2264's age are reported in the literature; they
lie between 3  and 10 million years (e.g., Sung et al. \cite{sun04}, Sagar et al. \cite{sag86}). If stars are still being formed in NGC\,2264, some of its members might even be only 0.15 million years old  (Sung et al. \cite{sun97}).

NGC 2264 is not significantly affected by reddening and differential reddening is negligible as well. The mean value of \mbox{$E(B-V) = 0.071 \pm 0.033$ mag} reported by Sung et al. (\cite{sun97}) is in good agreement with values found by other authors (e.g., Walker \cite{wal56}, Park et al. \cite{par00}).

The first ever discovered PMS pulsators, the two PMS \dsct\ stars V\,588\,Mon and V\,589\,Mon (Breger et al. \cite{bre72}), are members of NGC\,2264. Since then several studies have been already devoted to the search for and analysis of its pulsating members, e.g., using data from the MOST space telescope (Zwintz et al. \cite{zwi09}). 

Using the high-precision time series photometry of all accessible stars in the field of NGC\,2264 obtained with the CoRoT satellite, we  conducted a homogeneous survey for pulsating candidates in the magnitude range between 10 and 16\,mag in $R$. Cluster members showing variability on time scales between about 0.3 to a few days have been the prime candidates to search for PMS \gdor\ pulsation.
Two of these are NGC\,2264\,VAS\,20 (CoID0500007018) and NGC\,2264\,VAS\,87 (CoID0223979759), named VAS\,20 and VAS\,87 hereafter, and are subject of this work.

VAS\,20 (RA$_{2000}$ = 06:40:05.66, DE$_{2000}$ = +09:35:49.3) has a $V$ magnitude of 11.248\,mag (Sung et al. \cite{sun97}) and a photometrically determined spectral type F5 (Young \cite{you78}). An X-ray flux was detected by ROSAT/HRI observations (Flaccomio et al. \cite{fla00}) favoring the star's youth. Additionally, F{\H u}r{\'e}sz et al. (\cite{fur06}) listed the star as a cluster member according to its radial velocity. The membership probability is given as 94\% by Vasilevskis et al. (\cite{vas65}). 
King (\cite{king98}) reports an effective temperature (\Teff) of 6219 $\pm$ 140\,K using $B-V$ and $V-I$ photometry of Sung et al. (\cite{sun97}) and calibrations of Bessel (\cite{bes79}), and a Lithium abundance $\log\,n$(Li) = 3.56 derived from low-resolution spectroscopy.

VAS\,87 (RA$_{2000}$ = 06:40:56.95, DE$_{2000}$ = +09:48:40.7) has a $V$ magnitude of 12.323\,mag (Sung et al. \cite{sun97}) and a photometrically determined spectral type F9 (Young \cite{you78}). The membership probability is given as 96\% by Vasilevskis et al. (\cite{vas65}). VAS\,87 is also an X-ray source detected by the CHANDRA satellite (Ramirez et al. \cite{ram04}). Therefore the star is very likely to be young and a member of NGC\,2264. In the study by King (\cite{king98}) a \Teff = 6135 $\pm$ 140\,K and a $\log\,n$(Li) = 3.65 are given.

\section{The CoRoT light curves}
\subsection{Observations and data reduction}
The CoRoT satellite (Baglin \cite{bag06}) was launched on 2006, Dec 27th, from Baikonur aboard a Soyuz rocket into a polar, inertial circular orbit at an altitude of 896\,km. 
CoRoT carries a 27-cm telescope and can observe stars inside two cones of 10$^{\circ}$ radius, one at $RA=06:50$ and the other at $RA=18:50$. The field of view of the telescope is almost circular with a diameter of 3.8\,$^{\circ}$. Each CCD is a square of 1.3\,$^{\circ}$. The filter bandwidth ranges from 370 to 1000\,nm. 

CoRoT observed NGC\,2264 including VAS\,20 and VAS\,87 for the first time for 23.4 days in March 2008 during the Short Run SRa01 within the framework of the {\it Additional Programme} (Weiss \cite{wei06}). A second short run, SRa05, on the cluster NGC\,2264 was conducted in December 2011 / January 2012 during which only VAS\,87 was re-observed with a time base of about 39 days.

For both observing runs, the complete cluster was placed in one Exofield CCD and data were taken for all stars in the accessible magnitude range, i.e., from 10 to 16\,mag in $R$. The 100 brightest stars in the field of NGC\,2264 were the primary targets to search for stellar pulsations among PMS cluster members.

\begin{figure}[htb]
\centering
\includegraphics[width=0.48\textwidth,clip]{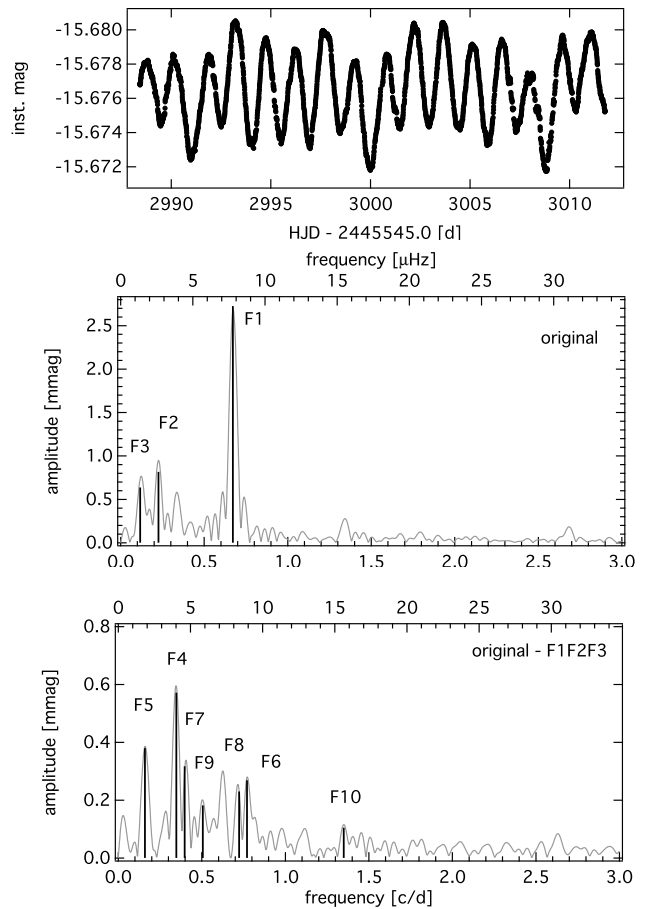}
\caption{CoRoT light curve (top panel) for VAS\,20, original amplitude spectrum (grey) from 0 to 3\,\cd\  (bottom X axis), i.e., from 0 to 34.72\,$\mu$Hz (top X axis), with the first three intrinsic frequencies identified (black), residual amplitude spectra after prewhitening the three highest amplitude frequencies (where the remaining 7 intrinsic frequencies are again marked in black). Note the different y-axes scales.}
\label{vas20}
\end{figure}

The reduced N2 data for both stars and both runs were extracted from the CoRoT data archive. The CoRoT data reduction pipeline (Auvergne et al. \cite{auv09}) flags those data points that were obtained during passages of the satellite over the South Atlantic Anomaly (SAA) and replaces them with linearly interpolated values. We did not use these `SAA-flagged' data points in our analysis. The light curves of VAS\,20 and VAS\,87 obtained in 2008 (see Figures \ref{vas20} and top left panel in Figure \ref{vas87}) consist of 2704 and 2600 data points, respectively, with a sampling time of 512 sec. The 2011/12 data set of VAS\,87 has been observed with a sampling time of 32 sec and therefore comprises 90073 data points. We checked for the presence of contaminant stars in the point-spread function and mask used for the observations for all three light curves and found no contaminating sources.

For VAS\,20 chromatic light curves in red, green and blue were obtained by CoRoT, while for VAS\,87 data were taken only in white light. The chromatic light curves were used to check the variability of VAS\,20. The variations are present in all three colors with the expected different photometric amplitudes. The top panels of Figures~\ref{vas20} and \ref{vas87} show the light curves in white light for the two stars where the fluxes obtained in the three colors for VAS\,20 were summed to increase the signal-to-noise. 

VAS\,20 was also observed by the MOST (Walker et al. \cite{wal03}) satellite in 2006 (Zwintz et al. \cite{zwi09}) and VAS\,87 was included in the MOST observations of NGC\,2264 in 2011/12. As the two stars are rather faint targets for MOST, the low amplitude variability is completely hidden in the noise and the respective light curves cannot be used for any further investigation.

\subsection{Comparison of the CoRoT data of SRa01 and SRa05} \label{corot}

\begin{figure*}[htb]
\centering
\includegraphics[width=0.9\textwidth,clip]{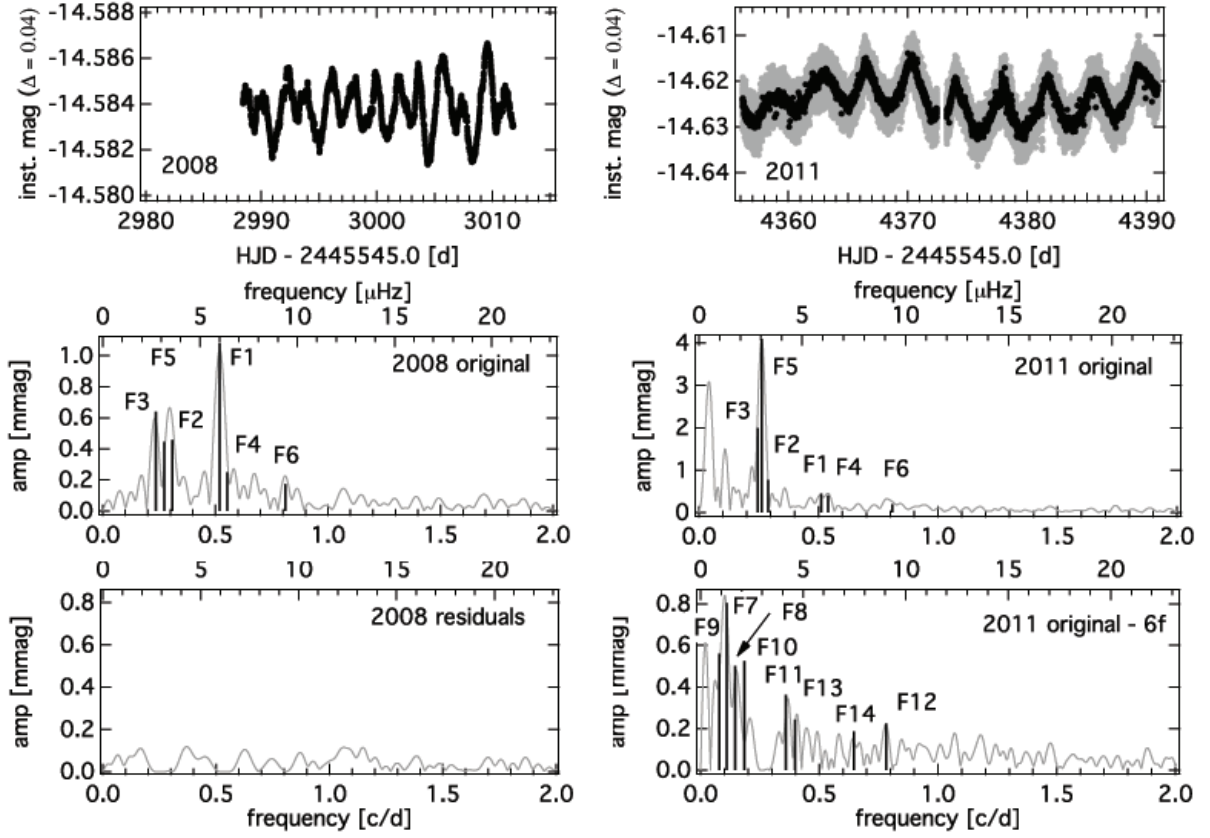}
\caption{CoRoT light curves (top panels) for VAS\,87 from SRa01 (2008 data, left panel) and SRa05 (2011/12 data, right panel; light curve with 32sec sampling in grey and binned to 512sec in black) to the same X-axes scales, original amplitude spectra (grey) from 0 to 2\,\cd\ (bottom X axis) i.e., from 0 to 23.15\,$\mu$Hz (top X axis), with the six common intrinsic frequencies identified in black (middle panels), residual amplitude spectra after prewhitening the six frequencies (bottom panels) where the additional eight frequencies found in the 2011/12 data are marked in black (bottom right panel). Note the different Y-axes scales in the top and middle panels.}
\label{vas87}
\end{figure*}

The light curves of VAS\,87 (Figure~\ref{vas87}) observed in 2008 (i.e., SRa01) and 2011/12 (i.e., SRa05) show a dramatic difference in the amplitudes of the observed variability. A detailed comparative study of the characteristics of the NGC\,2264 observations by CoRoT obtained in both years showed that several instrumental effects are present.
These instrumental effects, as described below, make it very difficult for us to interpret amplitude changes between the two observing years. For that reason we do not consider amplitude variations in this paper.

Different instrumental settings were used for CoRoT in 2008 and 2011/12 which might affect the measured amplitudes: 
For the observations in 2008, the CCD E1 was used, but as it is not operative any more, the observations in 2011/12 were conducted using the second Exoplanet CCD, E2.
For the short run SRa01 the best possible observing season for NGC\,2264 was chosen (i.e., March 2008) minimizing the influence of scattered light. Due to the coordinated simultaneous observations of CoRoT with the satellites Spitzer (Werner et al. \cite{wer04}), Chandra (Weisskopf et al. \cite{wei02}) and MOST (Walker et al. \cite{wal03}) in 2011/12 not the optimum observing period was taken (i.e., December 2011 and January 2012). 
Therefore different orientations of the CoRoT satellite in the two runs had to be chosen and different photometric templates had to be used. The different orientation of the cluster on the CCD could have some impact on the properties of the observations, e.g., the CCD temperature in the 2011/12 observations was higher than in 2008. 
Additionally, the aging of the CCD resulted in an increased rate of hot pixels and a higher dark current level. 

Many of these effects are corrected by the CoRoT pipe-line in a first approximation, and do not affect the time-domain properties.
But the situation is more difficult concerning the amplitudes of individual modes.
For instance, aperture photometry can be affected by the background levels of the stellar field and of the CCD.
An uncertainty on the estimation of the background level of 10\%  leads to an uncertainty on the amplitude of the same order.
The estimate of the background used by the standard correction in the CoRoT pipe-line depends on many factors, such as the proximity of the background windows, the orientation of the CCD, and the shape of the photometric template. So it can vary from one target to another.
A detailed study of these effects is currently in preparation (Baglin et al. \cite{bag12}). 

As the above described effects influence the ultra-precise CoRoT data significantly, the relative amplitudes for VAS\,87 obtained from the SRa01 data are different to those derived from the SRa05 data. 
This is why in this paper we do not discuss amplitude variations.

\subsection{Frequency Analysis}\label{frequencyanalysis}
For the frequency analyses, we used the software package Period04 (Lenz \& Breger \cite{len05}) that combines Fourier and least-squares algorithms. Frequencies were then prewhitened and considered to be significant if their amplitudes exceeded four times the local noise level in the amplitude spectrum (Breger et al. \cite{bre93}, Kuschnig et al. \cite{kus97}).
We verified the analysis using the SigSpec software (Reegen \cite{ree07}). SigSpec computes significance levels for amplitude spectra of time series with arbitrary time sampling. The probability density function of a given amplitude level is solved analytically and the solution includes dependences on the frequency and phase of the signal.

In 2008, VAS\,20 and VAS\,87 were observed with a sampling time of 512\,s, corresponding to a Nyquist frequency of 84\,\cd; hence, the frequency analysis was conducted in the range from 0 to 84\,\cd. The Rayleigh frequency resolution $1/T$ corresponds to 0.04\,\cd, where $T$ is the length of the observations (23.4\,days). 
In the case of two frequencies present within $1/T$, Kallinger et al. (\cite{kal08}) have shown using numerical simulations that the upper limit for the frequency error is $1/(4\cdot\,T)$ (see also their Figure 3) if the SigSpec significance ($sig$) is higher than 16. In our case this corresponds to a resolution of 0.01\cd.
For $sig\,<\,16$, the frequency error is $1/(T*\sqrt{sig})$, i.e., the values given in parentheses in Tables~\ref{VAS20-fs} and \ref{VAS87-fs}. 
In our case only F10 in VAS\,20 has $sig\,<\,16$.

In 2011, VAS\,87 was observed with a sampling time of 32\,s, corresponding to a Nyquist frequency of 1218\,\cd. The complete run lasted for about 39 days, but due to instrumental effects, the first few days had to be omitted from the analysis. The total length of the data set (shown in Figure \ref{vas87}) used in this investigation comprises $\sim$35 days, which corresponds to a Rayleigh frequency resolution of 0.028\,\cd. Applying the Kallinger et al. (\cite{kal08}) criterion, we get an upper limit of the frequency error of 0.007\,\cd, if the SigSpec significance is higher than 16. The frequency errors for $sig\,<\,16$ are given in Table~\ref{VAS87-fs}.

Using the CoRoT light curve from 2008, VAS\,20 shows ten intrinsic frequencies in the range from 0 to 1.35\,\cd\ (Figure~\ref{vas20} and Table~\ref{VAS20-fs}).  Linear combinations between the frequencies have been sought within three times the frequency uncertainty.
Three frequencies, i.e., F4, F9 and F10, might be explained as linear combinations, the other seven are clearly independent from each other. 
We believe that the frequency corresponding to the rotational period of VAS\,20 is F1 (see Section 5).

For VAS\,87 the data sets from 2008 and 2011 were analyzed independently and the results were then compared to each other. As the data set obtained in 2011 is significantly longer, the noise level in the amplitude spectrum is decreased and additional frequencies could be identified. Six frequencies are common to both data sets (see Table~\ref{VAS87-fs}), eight additional frequencies were found in the 2011 data set. From these 14 frequencies in total, seven could be explained as linear combinations (again within three times the frequency uncertainties) which are identified in the last column of Table~\ref{VAS87-fs} and seven are clearly independent from each other. 
Due to the instrumental differences in the observations from 2008 and 2011 (see Section \ref{corot}), the respective amplitudes of the common frequencies cannot be used for an astrophysical interpretation. 
For VAS\,87 we believe that F5 is the frequency which indicates the rotational period (see Section 5).

Applying the criteria by Kallinger et al. (\cite{kal08}), all frequencies for VAS\,20 and VAS\,87 are resolved and are therefore considered intrinsic to the stars.

\begin{table*}[htb]
\caption{Results of the frequency analysis of VAS\,20: frequencies and periods, amplitudes, signal-to-noise values and SigSpec significances are given sorted by the prewhitening sequence. The respective last-digit errors of the frequencies computed according to Kallinger et al. (\cite{kal08}) are given in parentheses. Possible linear combinations are listed in the last column.}
\label{VAS20-fs}
\begin{center}
\begin{tabular}{lllccrrl}
\hline
\multicolumn{1}{c}{No.} & \multicolumn{1}{c}{freq} & \multicolumn{1}{c}{freq} & \multicolumn{1}{c}{period} & \multicolumn{1}{c}{amp} & \multicolumn{1}{c}{S/N} & \multicolumn{1}{c}{sig}  & \multicolumn{1}{c}{combi}   \\
\multicolumn{1}{c}{\#} & \multicolumn{1}{c}{[\cd]}  & \multicolumn{1}{c}{[$\mu$Hz]} & \multicolumn{1}{c}{[d]} & \multicolumn{1}{c}{[mmag]} & \multicolumn{1}{c}{  }  & \multicolumn{1}{c}{  } & \multicolumn{1}{c}{ } \\
\hline
F1 & 0.672(2) & 7.78(2) & 1.488 & 2.726	 & 14.8 & 428.9 & \\
F2 & 0.226(3) & 2.61(4) & 4.427 & 0.818	 & 8.6 & 164.6 & \\
F3 & 0.117(4) & 1.35(5) & 8.566 & 0.638 & 7.4 & 118.6 &  \\
F4 & 0.347(4) & 4.02(4) & 2.879 & 0.572 & 8.9 & 136.3 & 3\,F3\\
F5 & 0.160(5) & 1.85(6) & 6.257 & 0.380	 & 6.5 & 72.8 & \\
F6 & 0.760(6) & 8.80(7) & 1.315 & 0.299	 & 4.7 & 47.4 & \\
F7 & 0.397(8) & 4.60(9) & 2.518 & 0.217	 & 4.3 & 27.2 & \\
F8 & 0.723(9) & 8.37(10) & 1.383 & 0.320 & 7.5 & 22.3 & \\
F9 & 0.506(9) & 5.86(11) & 1.977 & 0.182 & 4.5 & 20.3 & F1-F5 \\
F10 & 1.350(13) & 15.62(15) & 0.741 & 0.105 & 3.6 & 11.5 & 2 F1 \\
\hline
\end{tabular}
\end{center}
\end{table*}

\begin{table*}[htb]
\caption{Results of the frequency analysis of VAS\,87 using data from 2008 and 2011: frequencies and periods, amplitudes, signal-to-noise values and SigSpec significances for both years are given sorted first by the prewhitening sequence of the 2008 data and then of the 2011/12 data. The respective last-digit errors of the frequencies computed according to Kallinger et al. (\cite{kal08}) are given in parentheses. In the last column possible linear combinations are listed.}
\label{VAS87-fs}
\begin{center}
\begin{tabular}{lllcrrllcrrrl}
\hline
\multicolumn{1}{c}{No.} & \multicolumn{1}{c}{freq08} & \multicolumn{1}{c}{freq08}  & \multicolumn{1}{c}{amp08} & \multicolumn{1}{c}{S/N-08} & \multicolumn{1}{c}{sig08}   & 
\multicolumn{1}{c}{freq11} & \multicolumn{1}{c}{freq11} &  \multicolumn{1}{c}{amp11} & \multicolumn{1}{c}{S/N-11} & \multicolumn{1}{c}{sig11} & \multicolumn{1}{c}{period} & \multicolumn{1}{c}{combi} \\
\multicolumn{1}{c}{\#} & \multicolumn{1}{c}{[\cd]}  & \multicolumn{1}{c}{[$\mu$Hz]}  & \multicolumn{1}{c}{[mmag]} & \multicolumn{1}{c}{  }  & \multicolumn{1}{c}{  }  & 
 \multicolumn{1}{c}{[\cd]}  & \multicolumn{1}{c}{[$\mu$Hz]}  & \multicolumn{1}{c}{[mmag]} & \multicolumn{1}{c}{  }  & \multicolumn{1}{c}{  } & \multicolumn{1}{c}{[d]} & \multicolumn{1}{c}{  } \\
\hline
F1 & 0.519(3) & 6.01(3)  & 1.081 & 9.6 & 219.3 & 0.511(2) & 5.92(2) & 0.451 & 5.1 & 323.8 & 1.925 & \\
F2 & 0.309(3) & 3.58(4) & 0.410 & 10.6 & 167.6 & 0.291(3) & 3.36(3) & 0.769 & 13.8 & 100.1 & 3.236 &  \\
F3 & 0.236(4) & 2.73(5) &  0.641 & 9.0 & 114.1 & 0.251(1) & 2.90(1) & 1.996 & 19.9 & 748.1 & 4.235 & \\
F4 & 0.554(7) & 6.41(9) & 0.250 &  6.1 & 31.9 & 0.540(2) & 6.25(2) & 0.404 & 6.2 & 243.6 & 1.806 & F2+F3  \\
F5 & 0.274(9) & 3.17(10) & 0.447 & 9.6 & 23.1 & 0.2620(3) & 3.033(4) & 4.098 & 23.6 & 7980.9 & 3.655 & F1-F3 \\
F6 & 0.811(9) & 9.38 (55) & 0.174 & 4.1 & 19.7 & 0.809(3) & 9.36(4) & 1.943 & 4.06 & 73.0 & 1.233 & \\
F7 & & & & & &	0.110(1) & 1.27(1) & 0.804 & 7.1 & 916.5 & 9.091 & \\
F8  & & & & & &	0.145(2) & 1.68(2) & 0.501 & 5.3 & 307.8 & 6.897 & \\
F9  & & & & & &	0.079(2) & 0.92(2) & 0.560 & 7.1 & 365.3 & 12.658 & 2\,F2-2\,F3\\
F10  & & & & & &	0.184(2) & 2.13(2) &	0.526 & 7.1 & 235.3 & 5.435 & \\
F11 & & & & & &	0.359(2) & 4.15(2) & 0.362 & 5.9 & 211.1 & 2.786 & F3+F7\\
F12 & & & & & &	0.782(3) & 9.05(3) & 0.225 & 3.8 & 95.6 & 1.279 & 3\,F1-3\,F3\\
F13 & & & & & &	0.399(4) & 4.62(4) & 0.243 & 4.7 & 65.9 & 2.506 & F1-F7\\
F14 & & & & & &	0.645(4) & 7.47(5) & 0.186 & 4.1 & 54.3 & 1.550 & 3\,F3-F7\\
\hline
\end{tabular}
\end{center}
\end{table*}

\section{Spectroscopic analysis}\label{spectroscopy}

Fundamental parameters and abundances of both stars were determined from spectra obtained on 2010, December 15th and 16th with the Robert G. Tull Coud\'e Spectrograph (TS) on the 2.7-m telescope of Mc\,Donald Observatory. In the adopted configuration the cross-dispersed \'echelle spectrograph has a resolving power of 60\,000 and delivered spectra of VAS\,20 and VAS\,87 with signal-to-noise ratios (S/N) per pixel, calculated at $\sim$5000\,\AA, of 100 and 65, respectively. Each spectrum covered the wavelength range from 3633--10849\,\AA\ with gaps between the \'echelle orders at wavelengths larger than 5880\,\AA. 

Bias and Flat Field frames were obtained at the beginning of each night, while several Th-Ar comparison lamp spectra were obtained each night for wavelength calibration purposes. The reduction was performed using the Image Reduction and Analysis Facility\footnote{IRAF (http://iraf.noao.edu) is distributed by the National Optical Astronomy Observatory, which is operated by the Association of Universities for Research in Astronomy (AURA) under cooperative agreement with the National Science Foundation.} (IRAF). The spectra were normalized by fitting a low order polynomial to carefully selected continuum points. 

To compute model atmospheres of the target stars we employed the \llm\ stellar model atmosphere code by Shulyak et al (\cite{llm}), while we calculated synthetic spectra with \synth\  (Kochukhov \cite{synth3}). Our main source of atomic parameters for spectral lines is the \vald\ database (Kupka et al. \cite{vald2}). The LTE abundance analysis was based on equivalent widths, analyzed with a modified version (Tsymbal \cite{vadim}) of the {\sc WIDTH9} code (Kurucz \cite{kurucz1993a}). 

\begin{figure}[htb]
\centering
\includegraphics[width=0.45\textwidth,clip]{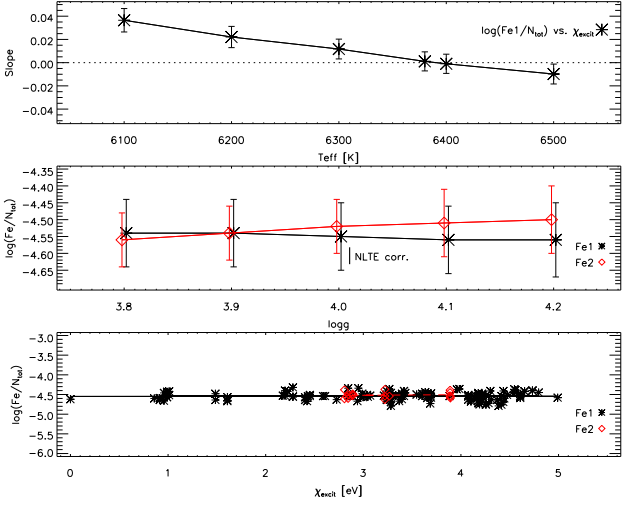}
\caption{Top panel: correlation between line abundance and excitation potential for \ion{Fe}{i} as a function of \Teff\ for VAS\,20. Middle panel: \ion{Fe}{i} (asterisks) and \ion{Fe}{ii} (diamonds) abundance as a function of \logg\ for VAS\,20. A small horizontal shift was applied to the points for visualisation purposes. Bottom panel: excitation equilibrium for both \ion{Fe}{i} (asterisks) and \ion{Fe}{ii} (diamonds) lines for VAS\,20. The correlations are shown respectively by a solid and a dashed line.}
\label{equilibria_vas20}
\end{figure}

\begin{figure}[htb]
\centering
\includegraphics[width=0.45\textwidth,clip]{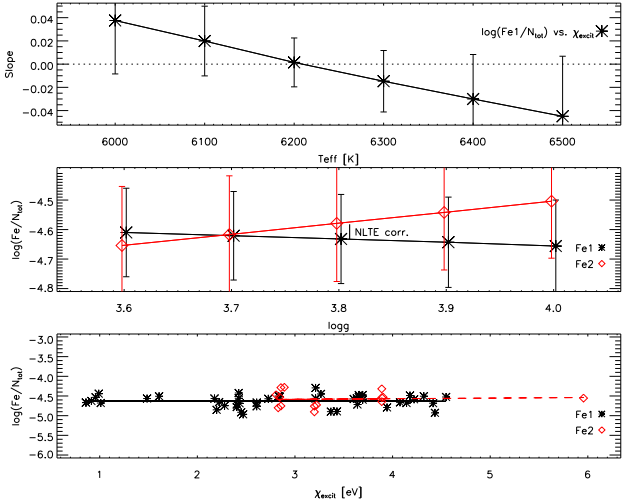}
\caption{Same as Figure~\ref{equilibria_vas20}, but for VAS\,87.}
\label{equilibria}
\end{figure}

We determined the microturbulence velocity (\vmic) imposing the equilibrium between line abundance and equivalent width for \ion{Fe}{i} lines, \logg\ imposing the ionisation equilibrium for Fe, and \Teff\ imposing the excitation equilibrium for \ion{Fe}{i} lines. For VAS\,87 we obtained \Teff=6220$\pm$150\,K, \logg=3.8$\pm$0.2 and \vmic=1.7$\pm$0.5\,\kms, while for VAS\,20 we obtained \Teff=6380$\pm$150\,K, \logg=4.0$\pm$0.2 and \vmic=1.8$\pm$0.5\,\kms. 
The excitation and ionisation equilibria obtained for VAS\,20 and VAS\,87 are shown in Figures~\ref{equilibria_vas20} and \ref{equilibria}. For both stars, the measured equivalent widths span between about 20 and 250\,m\AA. Given the fact that the uncertainty on \Teff\ is only statistical and does not take into account the possible presence of various systematic effects, we increased the uncertainty to 200\,K. By imposing the ionisation equilibrium, we took into account NLTE effects for Fe, as given by Mashonkina et al. (\cite{mash11}). 

It is extremely important to check the parameters obtained in this way with other indicators as this method might be strongly affected by systematics (see for example Ryabchikova et al. \cite{tanya09} and Fossati et al. \cite{fos11}). For this reason we used hydrogen lines and lines with extended wings to check \Teff\ and \logg, respectively. Note that at this temperaure the hydrogen lines have the advantage of not being very sensitive to gravity variations.

In PMS stars various spectral lines, particularly the hydrogen lines, could show emission as young stars might still be surrounded by the remnants of their birth clouds. If the hydrogen lines would be partially or completely filled, the determination of the effective temperature  (\Teff) from these lines would be impossible. For both VAS\,20 and VAS\,87 there is no sign of emission and if any it would affect just the line cores which cannot anyway be taken into account due to NLTE effects. We therefore used the H$\gamma$ line to estimate \Teff. H$\alpha$ was not covered and H$\beta$ is affected by a defect in the imaging system of the spectrograph. We normalized the H$\gamma$ line continuum using the artificial flat-fielding technique described in Barklem et al. (\cite{bar02}), which has proven to be successful with TS data (e.g., Fossati et al. \cite{fos11}). To reduce systematic uncertainties in the temperature determination due to the hydrogen lines normalization, we obtained low resolution ($R\sim$10000) spectra, covering H$\alpha$ (6470--6710\,\AA), with the 1.8-m telescope of the Dominion Astrophysical Observatory (Canada). Low resolution spectra allow a better control of the normalization. 

\begin{figure}[htb]
\centering
\includegraphics[width=0.45\textwidth,clip]{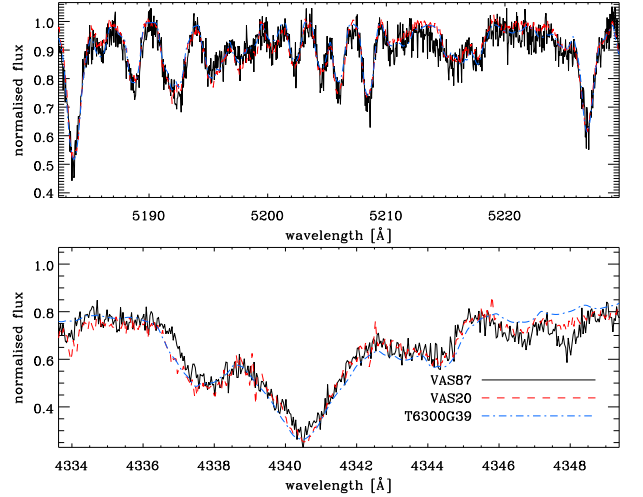}
\caption{Comparison between the observed spectra of VAS\,20 (dashed line) and VAS\,87 (solid line) in the region around the \ion{Mg}{i}b triplet (top panel) and H$\gamma$ line (bottom panel). Both panels present for comparison a synthetic spectrum (dash-dotted line) calculated with \Teff=6300\,K and \logg=3.9. To better appreciate the similarities between the two stars, we convolved the spectrum of VAS\,87 to the same \vsini\ of VAS\,20.}
\label{comparison}
\end{figure}

Within the uncertainties, the hydrogen lines confirmed the effective temperatures we derived for both stars with the excitation equilibrium. We confirmed the derived \logg\ values with the analysis of the \ion{Mg}{i} line profile at $\lambda$\,5172\,\AA\  (for more details on the adopted procedure see Fossati et al. \cite{fos11}), which, in this temperature regime, is particularly sensitive to gravity variations. Our fundamental parameters provide the best overall description of the available observables: hydrogen and metallic line profiles.
Our analysis shows that the two stars are very similar and this is demonstrated also in Figure~\ref{comparison} which shows a comparison between the spectra of the two stars.

By fitting synthetic spectra to several weakly blended lines we measured a \vsini\ of 42$\pm$2\,\kms\, for VAS\,20, and 18$\pm$1\,\kms for VAS\,87. 
Given the relatively large \vsini, we also performed the analysis of VAS\,20 using spectral synthesis (for more details see Fossati et al. \cite{fos07}), obtaining results consistent with the equivalent width results.
Given the \Teff\ of the two stars we expect the presence of some macroturbulence velocity (\vmac). For VAS\,20, \vsini\, is so large that the line profiles do not require any \vmac, while for VAS\,87, to be able to fit simultaneously the core and the wings of strong lines, we need a \vmac\, of about 12\,\kms. This value is slightly larger than what is expected for main sequence stars of this \Teff\  (Valenti \& Fisher \cite{valenti}), which might be connected to the fact that this star is in its PMS phase. For VAS\,20 we were not able to measure \vmac\, because of the rather large \vsini\, which prevented us from seeing the effects of \vmac\, broadening on the line wings. In this respect, the \vsini\, we obtained for VAS\,20 is likely to contain also the hidden information on \vmac, which cannot be directly measured.

\begin{table}[htb]
\caption[ ]{LTE atmospheric abundances in program stars with the error estimates based on the internal scattering from the number of analysed lines, $n$. For comparison purpose, the last column gives the abundances of the solar atmosphere calculated by Asplund et al. (\cite{asp09}).}
\label{abundance}
\begin{center}
\begin{tabular}{l|cc|cc|c}

\hline
Ion &\multicolumn{2}{|c|}{VAS\,20} &\multicolumn{2}{c|}{VAS\,87}  &  Sun \\                                  
    &$\log (N/N_{\rm tot})$ & $n$  &$\log (N/N_{\rm tot})$ & $n$ &$\log (N/N_{\rm tot})$  \\       
\hline
\ion{Li}{i }  & ~~$-$8.55$\pm$0.16 &  1 & ~~$-$8.60$\pm$0.10 &  1 & ~~$-$10.99~ \\      
\ion{Na}{i}   & ~~$-$5.83$\pm$0.08 &  2 & ~~$-$5.68:         &  2 & ~~$-$5.87~ \\	  
\ion{Mg}{i}   & ~~$-$4.63$\pm$0.03 &  4 & ~~$-$4.55$\pm$0.12 &  4 & ~~$-$4.44~ \\			   
\ion{Si}{i}   & ~~$-$4.58$\pm$0.09 &  7 & ~~$-$4.87$\pm$0.14 &  5 & ~~$-$4.53~ \\			    
\ion{Si}{ii}  & ~~$-$4.49:         &  1 & ~~                 &    & ~~$-$4.53~ \\			   
\ion{Ca}{i}   & ~~$-$5.64$\pm$0.09 & 16 & ~~$-$5.71$\pm$0.14 & 10 & ~~$-$5.70~ \\			   
\ion{Sc}{ii}  & ~~$-$8.97$\pm$0.09 &  6 & ~~$-$9.16:         &  1 & ~~$-$8.89~ \\			   
\ion{Ti}{i}   & ~~$-$7.00$\pm$0.08 &  6 & ~~$-$7.08$\pm$0.15 &  3 & ~~$-$7.09~ \\			   
\ion{Ti}{ii}  & ~~$-$7.03$\pm$0.06 &  8 & ~~$-$7.14$\pm$0.13 &  6 & ~~$-$7.09~ \\			   
\ion{Cr}{i}   & ~~$-$6.37$\pm$0.17 & 15 & ~~$-$6.39$\pm$0.10 &  6 & ~~$-$6.40~ \\			   
\ion{Cr}{ii}  & ~~$-$6.44$\pm$0.14 &  8 & ~~$-$6.06$\pm$0.08 &  5 & ~~$-$6.40~ \\			   
\ion{Mn}{i}   & ~~$-$6.62$\pm$0.15 &  4 & ~~$-$6.44$\pm$0.14 &  5 & ~~$-$6.61~ \\			   
\ion{Fe}{i}   & ~~$-$4.55$\pm$0.10 &127 & ~~$-$4.63$\pm$0.15 & 45 & ~~$-$4.54~ \\			   
\ion{Fe}{ii}  & ~~$-$4.52$\pm$0.08 & 17 & ~~$-$4.58$\pm$0.20 & 14 & ~~$-$4.54~ \\			   
\ion{Co}{i}   &                    &    & ~~$-$7.13:         &  1 & ~~$-$7.05~ \\			   
\ion{Ni}{i}   & ~~$-$5.77$\pm$0.10 & 16 & ~~$-$5.83$\pm$0.08 &  8 & ~~$-$5.82~ \\			   
\ion{Cu}{i}   & ~~$-$8.03:         &  1 & ~~$-$7.68:         &  1 & ~~$-$7.85~ \\			   
\ion{Y}{ii}   & ~~$-$9.94$\pm$0.08 &  2 & ~~$-$9.66$\pm$0.03 &  2 & ~~$-$9.83~ \\			   
\ion{Ba}{ii}  & ~~$-$9.29$\pm$0.04 &  3 & ~~$-$9.51$\pm$0.05 &  2 & ~~$-$9.86~ \\			   
\hline											     %
\hline											\end{tabular}
\end{center}
\end{table}

\begin{figure}[htb]
\centering
\includegraphics[width=0.45\textwidth,clip]{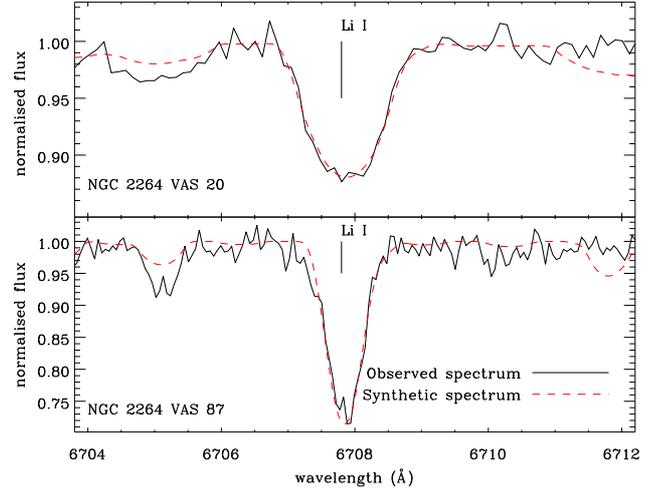}
\caption{Comparison between the observed spectra (black solid line) of VAS\,20 (upper panel) and VAS\,87 (lower panel) in the region of Li line at $\lambda\sim$6707\,\AA\, and synthetic spectra (red dashed line) calculated with the best fitting Li abundance. Note the different y-axes scales.}
\label{fig:lithium}
\end{figure}

For each star we determined the abundances of about 15 elements, which are given in Table~\ref{abundance}. Notice that the spectrum of VAS\,87 is badly affected by spikes, making both the continuum normalisation and the equivalent width measurements rather uncertain, in particular in the blue spectral region. This is responsible for the slightly larger abundance uncertainties for VAS\,87, compared to that of VAS\,20. For both stars, we obtained abundance values comparable to the solar ones, except for Ba and Li which are overabundant, in a reasonable agreement within the errors to the overall cluster metallicity of -0.15 (Lynga \cite{lynga}).
For Li we derived an abundance of $\log\,n$(Li)=3.49$\pm$0.16 for VAS\,20 and $\log\,n$(Li)=3.44$\pm$0.10 for VAS\,87 (see Figure~\ref{fig:lithium}). The difference in the uncertainties of the Li abundance for the two stars reflects the difference in the projected rotational velocity (\vsini).  We obtained the Li abundance from the lines at $\lambda\sim$6707\,\AA\ adopting hyperfine structure from Smith et al. (\cite{SLN}) and the meteoritic/terrestrial isotopic ratio Li$^6$/Li$^7$\,=\,0.08 (Rosman \& Taylor \cite{ros98}). The given errors on the Li abundance were calculated taking into account the uncertainties on the fundamental parameters (see Fossati et al. \cite{fos09}). 

\subsection{Spectral energy distribution and Hertzsprung-Russell diagram}
\begin{figure}[htb]
\centering
\includegraphics[width=0.45\textwidth,clip]{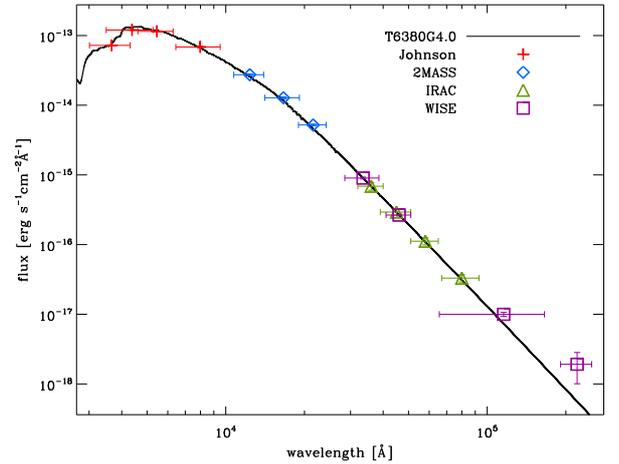}
\caption{Comparison between \llm\ theoretical fluxes (full line), calculated with the fundamental parameters derived for VAS\,20, with Johnson (crosses), 2MASS (diamonds), Spitzer (triangles), and WISE (squares) photometry converted to physical units.}
\label{sed_vas20}
\end{figure}

\begin{figure}[htb]
\centering
\includegraphics[width=0.45\textwidth,clip]{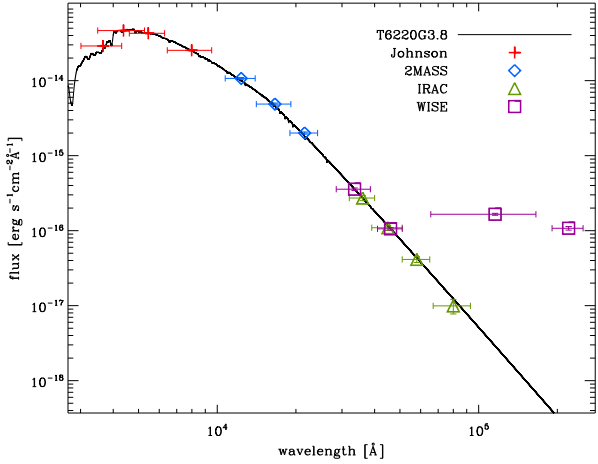}
\caption{Same as Figure~\ref{sed_vas20}, but for VAS\,87.}
\label{sed_vas87}
\end{figure}

Figure~\ref{sed_vas20} and \ref{sed_vas87} show the fits of the synthetic fluxes, calculated with the fundamental parameters derived for VAS\,20 and VAS\,87, respectively, to the observed Johnson (Sung et al. \cite{sun97}), 2MASS (Zacharias et al. \cite{zac05}), Spitzer (IRAC - Sung et al. \cite{sun09}) and WISE (Cutri et al. \cite{cut12}) photometry, converted to physical units. For this we used the calibrations provided by Bessel et al. (\cite{bes98}), van~der~Bliek et al. (\cite{van96}), Sung et al. (\cite{sun09}), and Wright et al. (\cite{wri10}), respectively. Adopting the cluster distance and reddening given by Sung et al. (\cite{sun97}), we estimated the stellar radius of VAS\,20 to be 3.5$\pm$0.5\,R/R$_\odot$, while for VAS\,87 we derived a radius of 2.2$\pm$0.5\,R/R$_\odot$. This comparison also confirms the temperature we obtained for the two stars.


Along the line of sight of VAS\,87 there is a dense cloud of material which we believe being behind the star, as it would otherwise be clearly visible in the spectral energy distribution, which we can fit with the same reddening applied for VAS\,20. 
This is also supported by the presence of only small amounts of reddening across the cluster reported in the literature (e.g., Sung et al. \cite{sun97}).
Figure~\ref{sed_vas87} shows also a steep increase in flux in the WISE photometry of the two red channels, but this feature is not present in the Spitzer photometry. It is therefore possible that the WISE photometry has been contaminated by the background cloud which would be rather bright at infrared wavelengths.

Using the spectroscopic temperatures, we put the two stars in the HR-diagram. We calculated the luminosity of each star photometrically, adopting the magnitudes in the $V$ band, a cluster distance of 759$\pm$83\,pc and a reddening of $E(B-V)$=0.071$\pm$0.033 (Sung et al. \cite{sun97}). We used the bolometric correction by Balona (\cite{bal94}). For VAS\,20 and VAS\,87 we obtained \logl=1.235 and 0.805, respectively. Given all the uncertainties and the fact that we adopted a bolometric correction valid for main-sequence stars, we decided to apply a conservative uncertainty on the luminosity of 0.1\,dex.
\begin{figure}[htb]
\centering
\includegraphics[width=0.45\textwidth,clip]{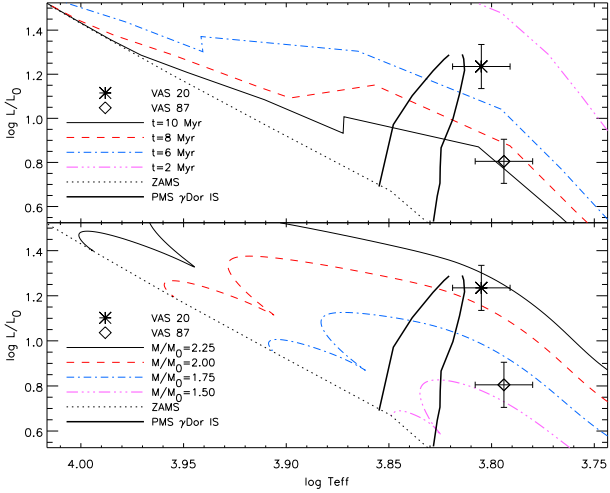}
\caption{Positions of VAS\,20 (asterisk) and VAS\,87 (diamond) in the HR-diagram in comparison with evolutionary tracks (bottom panel) and isochrones (top panel). In both panels the ZAMS is indicated by a dotted line. Evolutionary tracks for PMS stars with masses of 2.25\,M/M$_\odot$, 2.00\,M/M$_\odot$, 1.75\,M/M$_\odot$, and 1.50\,M/M$_\odot$ are shown respectively by a solid, dashed, dash-dotted and dash-triple dotted line. Isochrones with ages of 10\,Myr, 8\,Myr, 6\,Myr, and 2\,Myr are shown respectively by a solid, dashed, dash-dotted and dash-triple-dotted line. The borders of the \gdor\ instability strip (Dupret et al. \cite{dup04})  are marked with thick solid lines. }
\label{hr}
\end{figure}

Figure~\ref{hr} shows the position of the two stars in the HR-diagram, in comparison to PMS evolutionary tracks from Guenther (\cite{gue12}, priv. communication) using the YREC evolution code (Demarque et al. \cite{dem08}) with physics as described in Guenther et al. (\cite{gue09}) and isochrones by Demarque et al. (\cite{dem08}). 
For both evolutionary tracks and isochrones we adopted solar metallicity.
Also included in Figure~\ref{hr}  is the instability strip for \gdor\ stars (Dupret et al. \cite{dup04}).

The derived \Teff\ values are relatively low compared to the theoretical computations of the instability strip for PMS $\gamma$ Doradus stars (see Bouabid et al. \cite{bou11}, their Figure 8). The cool border of the predicted theoretical (PMS) $\gamma$ Doradus instability strip lies at about 6700\,K, but such a high \Teff\ would not allow to fit both hydrogen and metallic lines for both stars. Therefore the two stars lie slightly outside the predicted, theoretical PMS $\gamma$ Doradus instability strip (Bouabid et al. \cite{bou11}). 
This is not too surprising as similar discrepancies are found in other groups of (post-)main sequence pulsators. In particular several of the \gdor\ stars found with data from the {\it Kepler} satellite (Koch et al. \cite{koch10}) are located at temperatures similar to that of VAS\,20 and VAS\,87 (Balona et al. \cite{bal12}).

Additionally, the theoretical seismic properties of PMS $\gamma$ Doradus have been investigated without a confirmed member of this group of pulsators. A comparison of the observations to the PMS $\gamma$ Doradus theory might yield new constraints on the location of the respective instability region.
More theoretical and observational work will be performed in the future to investigate this in detail.
As the PMS \gdor\ instability strip coincides well with the instability strip of the (post-) main sequence \gdor\ stars, in Figure \ref{hr} we show the location of VAS\,20 and VAS\,87 in the HR-diagram with respect to the ``classical" \gdor\ instability strip by Dupret et al. (\cite{dup04}).

On the basis of the evolutionary tracks we derived masses and radii for the two stars. For VAS\,20 we obtained a mass of 2.17$\pm$0.15\,M$_\odot$, a radius of 3.5$\pm$0.5\,R$_\odot$ and a surface gravity of 3.69$\pm$0.11, while for VAS\,87 we obtained a mass of 1.55$\pm$0.15\,M$_\odot$, a radius of 2.1$\pm$0.5\,R$_\odot$ and a surface gravity of 3.98$\pm$0.11. The stellar radii, derived from the evolutionary tracks are in remarkable good agreement with what we obtained from the fit of the spectral energy distribution. 
Within one-sigma, the spectroscopic \logg\ values are also in agreement with those derived from the evolutionary tracks. 

VAS\,20 and VAS\,87 are two stars born from the same molecular cloud that are members of NGC\,2264 and that have nearly the same \Teff\ and \logg\ values, but slightly different masses, radii and luminosities (see Table \ref{pars} for an overview of the fundamental parameters) which place them at different locations in the HR-diagram. 
At present we can only speculate about the reason for this discrepancy.

One possible explanation might be connected to the presence of sequential star formation in NGC\,2264 and that its star forming activity is still going on (e.g., Flaccomio et al. \cite{fla99}). The ages of stars in the cluster lie between 0.8 and 8 million years where intermediate mass stars and very low mass stars were formed first and and massive stars down to about one solar mass were formed later on (Sung et al. \cite{sun97}). 
To assess if an age spread in NGC\,2264 would be a possible reason for the slight differences in the fundamental parameters of the two stars, we included the isochrones in Figure \ref{hr} and obtained an age of 4$\pm$1\,Myr for VAS\,20 and 10$\pm$2\,Myr for VAS\,87. In the present case, the higher mass star (VAS\,20) would have to be formed later than the lower mass star (VAS\,87).  This would fit to the theory of star formation in NGC\,2264 (Sung et al. \cite{sun97}) as described above. 

Another possible explanation for the differences in the two stars' fundamental parameters could be that VAS\,20 is part of a binary system. We have assessed that the CoRoT photometry is not affected by a contaminating source and we do not find any evidence for binarity in our spectroscopic measurements and in the spectral energy distribution. The \logg\ value of VAS\,87 derived from our spectroscopic data agrees well to the corresponding value determined from the evolutionary tracks. 
But for VAS\,20 there is a difference of about 0.3 dex between the two values. We therefore hypothesize that the luminosity of VAS\,20 is overestimated when extracted from spectroscopy because VAS\,20 might be part of an unresolved binary system.

\begin{table}[htb]
\caption[ ]{Fundamental parameters for VAS\,20 and VAS\,87.}
\label{pars}
\begin{center}
\begin{tabular}{lcc}
\hline
  &\multicolumn{1}{c}{VAS\,20} &\multicolumn{1}{c}{VAS\,87}  \\                                  
\hline
\Teff [K] & 6380 $\pm$ 150 & 6220 $\pm$ 150\, \\
\logg & 4.0 $\pm$ 0.2 & 3.8 $\pm$ 0.2 \\
\vmic\, [km s$^{-1}$] & 1.8 $\pm$ 0.5 & 1.7 $\pm$ 0.5 \\
\vsini\, [km s$^{-1}$] & 42 $\pm$ 2 & 18 $\pm$ 1 \\
$M$ [\Msun] &  2.17 $\pm$ 0.15 & 1.55 $\pm$ 0.15 \\
log $L$ [\Lsun] & 1.2 $\pm$ 0.1 & 0.8 $\pm$ 0.1 \\
$R$ [\Rsun] & 3.5 $\pm$ 0.5 & 2.2 $\pm$ 0.5 \\
\hline											
\end{tabular}
\end{center}
\end{table}

\subsection{Comparison with photometric colors}
Young stars are often still surrounded by the remnants of their birth clouds which affects the observed photometric colors. Therefore the application of photometric calibrations (developed for main sequence stars) to PMS objects will often lead to erroneous values. 

In the present case, we used the available Str\"omgren photometry (Strom et al. \cite{str71}, Neri et al. \cite{neri93}) and the photometric calibration by Moon \& Dworetsky (\cite{moon}) to test this hypothesis. For VAS\,20 we get \Teff=5800\,K, \logg=3.6 and a metallicity [Fe/H] = -0.76 and \Teff=6400\,K, \logg=4.6 and a metallicity [Fe/H] = -0.61 for VAS\,87. Although these \Teff\ values might be used as a first estimate, the \logg\ values of 3.6 and 4.6 are, respectively, too low and too high for these stars and the quite low [Fe/H] values do not match the overall cluster metallicity of -0.15 (Lynga \cite{lynga}). This illustrates that the results of photometric calibrations for the determination of the fundamental parameters of PMS stars cannot be trusted and that only spectroscopic measurements can give reliable values.

\section{Rotation \& Pulsation}
  
As star spots and \gdor\ pulsation cause variability on the same time scales, i.e., in the low frequency domain, we here discuss the nature of the observed variations in VAS\,20 and VAS\,87 which we attribute to a combination of rotation and pulsation. 

Surface inhomogeneities can cause complex variability features in the  light curves and amplitude spectra due to multiple spots on the surface, differential rotation or migrating star spots (Strassmeier \cite{str09}). Nevertheless, frequencies originating from rotational modulation are mostly accompanied by non-linear effects which can be identified in the amplitude spectra as harmonics and linear combinations. 

Potential PMS \gdor\ pulsators are located in the same region of the HR-diagram as the T Tauri stars which are known to show regular, semi-regular and irregular variabilities. For the classical T Tauri stars in NGC\,2264, Alencar et al. (\cite{ale10}) report a mean rotational period between 3 and 4 days using data obtained during the same CoRoT run in 2008. 

\begin{figure}[htb]
\centering
\includegraphics[width=0.45\textwidth,clip]{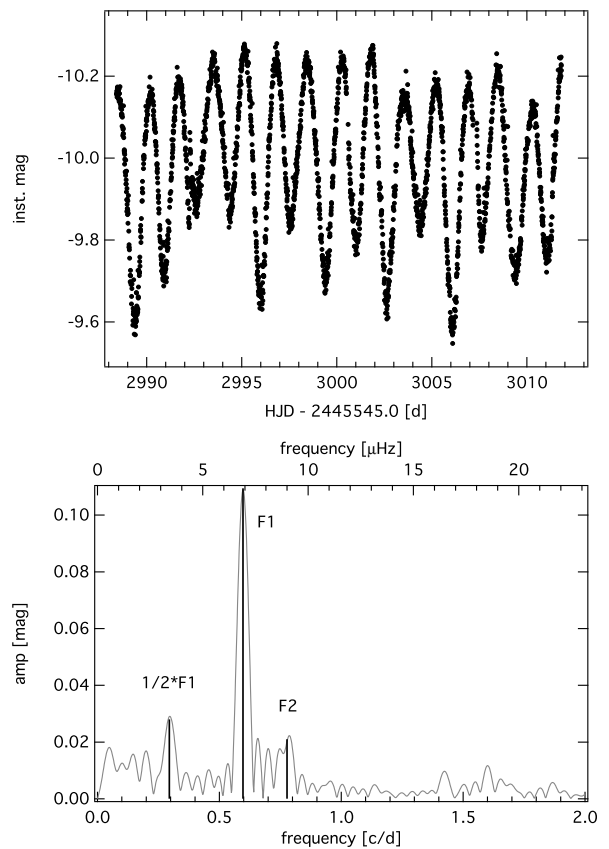}
\caption{The ``comparison'' T Tauri star CoID0223980447: CoRoT light curve (top panel) from SRa01 (2008) and original amplitude spectrum (bottom panel) from 0 to 2\,\cd\ (bottom X axis), i.e., from 0 to 23.15\,$\mu$Hz (top X axis), where the three significant frequencies are identified in black.}
\label{TTauri}
\end{figure}

Based on this information, we
performed additional frequency analyses of cluster members with known variability caused by spots (i.e., the classical T Tauri stars) which have spectral types from K0 to M0. These ``comparison'' stars were observed with CoRoT in the same run in 2008, i.e., have the same data set lengths, but are much cooler than VAS\,20 and VAS\,87. The low temperatures of the ``comparison'' stars ensure that their variability is likely to be produced by rotational modulation, as no pulsation (other than solar-like oscillations which are in a different frequency range, i.e., larger than the Nyquist frequency of 84\,\cd) can be present, and that their level of activity is equal or higher than that of VAS\,20 and VAS\,87. 
A typical example is given in Figure~\ref{TTauri}: the light curve (top panel) and amplitude spectrum (bottom panel) of CoID0223980447 observed in 2008 is shown. CoID0223980447 has a spectral type of K6. The frequency analysis shows two independant frequencies (F1 and F2 in Figure~\ref{TTauri}) and a third frequency which is 1/2\,F1 indicating the presence of non-linear effects.
From the frequency analyses of the spotted stars, we can confirm the findings by Alencar et al. (\cite{ale10}) and we infer that rotational modulation can be responsible for a maximum of three frequencies in the 23-day long CoRoT data sets from 2008.

Using data from the {\it Kepler} satellite, Balona et al. (\cite{bal12}) find three types of stars with multiple peaks in the frequency domain of \gdor\ pulsators: stars with one or two dominant peaks with asymmetric light curves (their ASYM group), stars with two close frequencies with symmetric light curves (their SYM group) and stars showing multiple low frequencies (their MULT group). While stars in the MULT group are very likely \gdor\ pulsators, there might be a mix of spotted stars and pulsators in their sample of the ASYM and SYM groups (Balona et al. \cite{bal12}). 
From their analyses the authors also conclude that the pulsation periods of their stars are close to the rotational periods.

The light curves of our two stars look rather symmetric, but once the peak with the highest amplitude (i.e., F1 for VAS\,20 and F5 for VAS\,87; see Figures \ref{vas20} and \ref{vas87}) is subtracted, multiple peaks at low amplitudes remain which strongly indicates the presence of pulsation.

Using a \vsini\ value of 42\,\kms\ and a stellar radius of 3.5\Rsun\ for VAS\,20, the maximum rotation period corresponds to $\sim$4.2\,d. We identify F1 at 0.672\,\cd, i.e., a period of about 1.5 days, to be likely caused by rotation because it is the highest and most dominant amplitude peak in our analysis with the other frequencies lying at significantly lower amplitudes (i.e., smaller than 1\,mmag). With these parameters, the projected inclination angle of the rotational axis is computed to be 23 degrees. 

For VAS\,87 a \vsini\ value of 18\kms with a stellar radius of 2.2\Rsun would correspond to a maximum rotation period of $\sim$6.1\,d. 
In our frequency analysis, the by far most dominant and highest amplitude peak in the 2011/12 data set is F5 at 0.262\cd corresponding to a period of 3.816\,d. In the 2008 data the highest amplitude peak is F1 at 0.519\,\cd which is about twice F5. As the surfaces of young stellar objects are likely not homogeneous, we assume that in 2008 the surface inhomogeneities were not that prominent than in 2011 and that pulsational and rotational activities occurred on similar levels in 2008. In this context, we still have to remember that the CoRoT photometry might be affected by instrumental effects to a presently unknown extent and that not all the large amplitude differences between the data sets from 2008 and 2011/12 might be interpreted astrophysically. 
We therefore conclude that F5 at 0.262\cd (value from the 2011 data set) corresponding to a period of 3.816\,d is likely to be caused by rotation. The corresponding projected inclination angle of the rotation axis is then 36 degrees for VAS\,87. 

We conclude that rotational modulation may explain only few of the detected frequencies in VAS\,20 and VAS\,87, but the clear multi-periodicity of both stars shows that rotation cannot be the only cause of their variability and that indeed $g$-modes have been observed.

\section{Conclusions}

VAS\,20 and VAS\,87 were observed for about 23.4 days with the CoRoT satellite during the short run SRa01. VAS\,87 was re-observed by CoRoT during the short run SRa05 in 2011/12 for about 39 days. The analyses of the high-precision photometric time series from space revealed ten and fourteen intrinsic frequencies between 0 and 1.5\,\cd\ which we attribute to be caused by a combination of rotation and $\gamma$ Doradus type pulsation. 
From high-resolution spectroscopy we determined the atmospheric parameters and chemical abundances for VAS\,20 and {\mbox VAS\,87}. 
Our  \Teff\ values of 6380\,K and 6220\,K agree, within the uncertainties, with those of King (\cite{king98}).

Using our measured \vsini\ values of 42\kms\ and 18\kms\ combined with the derived stellar radii of 3.5\Rsun\ and 2.2\Rsun\ for VAS\,20 and VAS\,87, respectively, we identify F1 at 0.672\cd\ for VAS\,20 and F5 at 0262\cd\ for VAS\,87 to be likely caused by rotation, while the other frequencies are interpreted as $g$-modes of \gdor\ pulsation.

In this context, we also compared VAS\,20 and VAS\,87 to several additional variable stars in NGC\,2264 in the spectral range from K0 to M0. At such cool temperatures variability can only be caused by spots on the stellar surface. From this analysis we find that rotational modulation can be responsible for three frequencies at maximum, but cannot explain the presence of ten and fourteen intrinsic frequencies identified in VAS\,20 and VAS\,87.

The probability that VAS\,20 and VAS\,87 are members of NGC\,2264 is very high due to their reported X-ray fluxes, radial velocity measurements and proper motions. The radial velocity (\vrad) of the cluster NGC\,2264 was derived from measurements of 13 stars and is given as 17.68\,$\pm$\,2.26\,\kms\ by Kharchenko et al. (\cite{kha05}). From our spectra we derive \vrad\ values of 17.7\,$\pm$\,1.0 and 19.3\,$\pm$\,0.9\,\kms\ for VAS\,20 and VAS\,87, respectively, which are in agreement with the cluster average. 

Our spectroscopic analyses add another crucial piece of information: the relatively high Li abundance we measured is in excellent agreement with the one obtained by Sestito \& Randich (\cite{sestito}) for other stars of similar \Teff\  in this cluster. The Li abundances reported by King (\cite{king98}) are higher than the values that we derived but agree within the uncertainties of 0.2 dex. These systematic differences might occur because we used a higher \Teff\ and took into account the full structure of the line. Additionally, our measurements of the equivalent widths are smaller by 20\,m\AA\ compared to King (\cite{king98}).

As the variability observed in VAS\,20 and VAS\,87 is very likely caused by a combination of rotational modulation and $g$-mode pulsation and as both types of variations occur on the same time scales, it is currently hard to unambiguously distinguish between them. But the multiple frequencies discovered in VAS\,20 and VAS\,87 provide a first strong indication for an observational detection of $\gamma$ Doradus pulsation in PMS stars which was theoretically predicted (Bouabid et al. \cite{bou11}).
These first observational candidates for PMS \gdor\ stars will be an important input for theory and help to continue the development of asteroseismic models for PMS stars.

\begin{acknowledgements}
KZ receives a Pegasus Marie Curie Post-doctoral Fellowship of the Research Foundation -- Flanders (FWO) and acknowledges support from an APART fellowship of the Austrian Academy of Sciences. 
We wish to thank the referee for useful comments that allowed us to clarify certain parts of this paper.
We thank Oleg Khochukov and Denis Shulyak for their support. We are grateful to Ettore Flaccomio for his extensive knowledge about NGC 2264 and the respective literature. We thank Michel Breger for fruitful discussions in preparation of the paper. Spectroscopic data were obtained with the 2.7-m telescope at Mc Donald Observatory, Texas, US and at the Dominion Astrophysical Observatory, Herzberg Institute of Astrophysics, National Research Council of Canada.
\end{acknowledgements}

\end{document}